\documentclass[a4paper,fleqn,usenatbib]{mnras}

\usepackage[T1]{fontenc}
\usepackage{ae,aecompl}

\usepackage{graphicx}	% Including figure files
\usepackage{amsmath}	% Advanced maths commands
\usepackage{amssymb}	% Extra maths symbols
\usepackage{longtable, multicol}
\usepackage{rotate}
\usepackage{lscape}
\usepackage{natbib}
\usepackage{latexsym,verbatim,xcolor}

\title[On the Formation of Deuterated Methyl Formate in Hot Corinos]{On the Formation of Deuterated Methyl Formate in Hot Corinos}

\author[Awad et al.]{
Zainab Awad$^{1}$\thanks{E-mail:zma@sci.cu.edu.eg}, Audrey Coutens$^{2,3}$, Serena Viti$^{4,5}$, and Jonathan Holdship$^{4,5}$\\
$^{1}$Department of Astronomy, Space Science, and Meteorology, Faculty of Science, Cairo University, 
Giza 11326, Egypt\\
$^{2}$Laboratoire d'Astrophysique de Bordeaux, Univ. Bordeaux, CNRS, B18N, all{\'e}e Geoffroy Saint-Hilaire, F-33615 Pessac, France\\
$^{3}$Institut de Recherche en Astrophysique et Plan{\'e}tologie, Universit{\'e} de Toulouse, UPS-OMP, CNRS, CNES, 9 av. du Colonel Roche, \\
31028 Toulouse Cedex 4, France\\
$^{4}$Leiden Observatory, Leiden University, PO Box 9513, 2300 RA Leiden, The Netherlands\\
$^{5}$Department of Physics and Astronomy, University College London, London WC1E 6BT, UK 
}

\date{Accepted XXX. Received YYY; in original form ZZZ}

\pubyear{2021}

\begin{document}
\label{firstpage}
\pagerange{\pageref{firstpage}--\pageref{lastpage}}
\maketitle

\begin{abstract}

Methyl formate, HCOOCH$_3$, and many of its isotopologues have been detected in astrophysical regions with considerable abundances. However, the recipe for the formation of this molecule and its isotopologues is not yet known. In this work, we attempt to investigate, theoretically, the successful recipe for the formation of interstellar HCOOCH$_3$ and its deuterated isotopologues. We used the gas-grain chemical model, UCLCHEM, to examine the possible routes of formation of methyl formate on grain surfaces and in the gas-phase in low-mass star-forming regions. Our models show that radical-radical association on grains are necessary to explain the observed abundance of DCOOCH$_3$ in the protostar IRAS~16293--2422. H-D substitution reactions on grains significantly enhance the abundances of HCOOCHD$_2$, DCOOCHD$_2$, and HCOOCD$_3$.
The observed abundance of HCOOCHD$_2$ in IRAS 16293--2422 can only be reproduced if H-D substitution reactions are taken into account.
However, HCOOCH$_2$D remain underestimated in all of our models.
The deuteration of methyl formate appears to be more complex than initially thought.  Additional studies, both experimentally and theoretically, are needed for a better understanding of the interstellar formation of these species. 
\end{abstract}
%%%%%%%%%%%%%%%%%%%%
\begin{keywords}
Astrochemistry -- stars: formation -- stars: low-mass -- ISM: abundances -- ISM: molecules.
\end{keywords}
%%%%%%%%%%%%%%%%%%%%%%%%%%%%%%%%%%%%%%%%%%%%%%%%%%%%
\section{Introduction}
\label{intro}
Complex organic molecules ({\it hereafter} COMs) are defined as interstellar molecules that are made of six or more atoms \citep{her09}, and 
accordingly, the simplest COM is methanol, CH$_3$OH. Several COMs have been detected in cores around massive and low-mass star forming regions 
(SFR) among which are  methyl formate (HCOOCH$_3$, {\it hereafter} MF), dimethyl ether (CH$_3$OCH$_3$, {\it hereafter} DME), and formamide 
(NH$_2$CHO) (e.g. \citealt{ike01,caz03,bot04b,jor05,bis07a,kah13}), as well as their isotopologues, in particular the deuterated ones, whether 
mono or multiply. 

Most deuterated COMs have been detected towards the low-mass protostar IRAS 16293--2422 A \& B (e.g. \citealt{cout16,jor16, jor18, man19, man20}) among which are those of MF ({\it hereafter} DMF). Using IRAM-30m and JCMT data from the TIMASSS program \citep{caux11}, \citet{dem10} reported tentatively the first detection of the mono-deuterated methyl formate DCOOCH$_3$ with a column density of 6 $\times$ 10$^{14}$ cm$^{-2}$ in both IRAS 16293A and IRAS 16293B where the column density of MF is estimated to be 1 $\times$ 10$^{16}$ cm$^{-2}$ and 9 $\times$ 10$^{15}$ cm$^{-2}$ in both cores, respectively. This data gives a deuterium fractionation ratio of around 6\% in IRAS 16293--2422 A \& B. 
The two singly deuterated forms HCOOCH$_2$D and DCOOCH$_3$ were later detected as part of the ALMA-PILS survey \citep{jor16} with D/H ratios of 1.7--2.8\% for the component A of IRAS 16293--2422 and 5.8--6.2\% for the component B (\citealt{jor18, man19}). 
More recently, \citet{man19} identified for the first time the doubly deuterated form HCOOCHD$_2$ using the same survey. The species was detected towards both cores A and B with a D/H ratio of 8 and 12\%, respectively.

Species larger than methanol were initially thought to form through chains of warm gas-phase chemistry triggered by the evaporation of mantle CH$_{3}$OH during the warming-up phase (T $\ge$ 100 K) of the star formation process (e.g. \citealt{gar06}). 
MF was long thought to be a product of the gas-phase reaction sequence, Eq. (\ref{eq:1}) below, which was used in several chemical models such as \citet{mil91}, \citet{cas93} and \citet{char95}. 
\begin{equation}
\label{eq:1}
\begin{split}
&\text{H$_2$CO + CH$_3$OH}^+_2  \longrightarrow \text{HCOOCH}^+_4 \\
&\text{HCOOCH$^+_4$ + e$^-$} \longrightarrow \text{HCOOCH}_3
\end{split}
\end{equation}
However, the first step of reaction (\ref{eq:1}), H$_{2}$CO + CH$_{3}$OH$^{+}_{2}$, was later proved to be highly inefficient because it possesses a  high activation energy of $\sim$ 15,000 K \citep{horn04}. 

Based on these findings, \citet{gar06} and \citet{gar08} constructed a gas-grain chemical model to study the formation of MF and other related COMs via surface reactions that involve the reactions of heavy radicals, Eq. (\ref{eq:2}), when they become mobile during the heating up of the medium by the new protostar. The authors found that the formation of HCOOCH$_3$ is efficient on grains through the reaction 
\begin{equation}
\label{eq:2}
\text{HCO + CH$_3$O} \longrightarrow \text{HCOOCH$_3$} 
\end{equation}
where both reactants are formed on grains through either the successive hydrogenation of CO (e.g. experiments by \citealt{wat02, wat03}) or the UV photo-dissociation of CH$_3$OH (e.g. experimental results by \citealt{bent07b, wat08}). 

\citet{mod10} showed that the formation of MF can occur, after cosmic ion irradiation, via the surface two-body reaction 
\begin{equation}
\label{eq:3}
\text{CO + CH$_3$OH} \longrightarrow \text{HCOOCH$_3$}.  
\end{equation}
The authors also calculated the reaction rate constant for Eq. (\ref{eq:3}). These experimental results (the reaction and the rate constant) were implemented in chemical models and proved their efficiency in the formation of methyl formate in the interstellar medium \citep{occ11}. 

In addition, \citet{vasy13a} used a full macroscopic Monte Carlo gas-grain model to explore the influence of the formation of grain icy mantle species in a multilayer (layer-by-layer) manner, during the early cold stages of the prestellar evolution, on the gas-phase chemistry triggered by the evaporation of icy mantle species, during the warming-up phase of the protostellar evolutionary stage. The model results showed that species that are essentially formed on grains such as HCOOCH$_3$ are affected more by the multilayer treatment than other species such as HCOOH and DME that are formed by gas-phase chemistry. The authors attributed their failure in reproducing the observed abundances of gaseous MF to the fact that its formation on grain surfaces in their multilayer simulations is severely affected by the limitation in the photodissociation reactions to the first four layers unlike normal two-phase models in which photochemistry occurs throughout the mantle.  After that, \citet{rud15} studied the formation of COMs in cold dense environments by introducing the combination between the Eley-Rideal mechanism and the induced COMs formation on grain surfaces. The findings showed that including these mechanisms into gas-grain models increases the abundances of MF and related species 10 to 100 times compared to models without these mechanisms. More recently, \citet{vasy17} found that adding the chemical or reactive desorption (RD) to the non-thermal desorption processes, into their gas-grain chemical model, successfully enhances the amount of gaseous COMs produced in the prestellar core L1544 to $\sim$ 10$^{-10}$ relative to the total H nuclei. These abundances depends on the efficiency of the RD mechanism, which in turn, rests on the icy composition of the first layer of the grain mantle where large abundances of CO and its hydrogenation products may reside. 

In addition to the formation of HCOOCH$_3$ on grain surfaces, gas-phase routes have been also investigated in chemical models. \citet {bal15} succeeded in reproducing the observed abundances of MF and DME in cold regions using pure gas-phase chemical models. In their models, the authors introduced a set of gas-phase reactions, Eq. (\ref{eq:4}), that forms MF from DME by converting the latter to the reactive intermediate radical CH$_3$OCH$_2$
\begin{equation}
\label{eq:4}
\begin{split}
&
\text{CH$_3$OH + OH } \longrightarrow \text{CH$_3$O + H$_2$O} \\
&
\text{CH$_3$O  + CH$_3$} \longrightarrow \text{CH$_3$OCH$_3$ + photon} \\
&
\text{CH$_3$OCH$_3$ + Cl (F)} \longrightarrow \text{CH$_3$OCH$_2$ + HCl (HF)} \\
&
\text{CH$_3$OCH$_2$ + O} \longrightarrow \text{HCOOCH$_3$ + H } 
\end{split}
\end{equation}
The results of \citet{bal15} models showed that gas-phase reactions are capable of reproducing the observed abundances of MF and DME in cold environments down-to 10 K simulating the cold dense prestellar phase. 

For the deuterated counterparts of methyl formate ({\it hereafter} DMF), \citet{dem10} suggested that DCOOCH$_3$ is formed on grain surfaces via radical-radical reactions similar to those forming HCOOCH$_3$, with HCO replaced by DCO. \citet{bent07b} experimental results suggested that the formation of DCO happens by the UV photo-dissociation of mantle CH$_3$OD. Other experiments suggested that H-D substitution reactions are important in the formation of partially deuterated methanol \citep{nag05}, and formaldehyde \citep{hid09}. Following the same approach partially deuterated HCOOCH$_3$ might result from H-D substitution of HCOOCH$_3$ during the ice phase change when these grains are warmed up by stellar radiation. \citet{taq14} investigated the formation and evolution of deuterated COMs including singly deuterated MF in low-mass star forming regions with a multilayer approach. A gas-grain astrochemical model was coupled with a one-dimensional dynamical model of a collapsing core. The model showed that, in the grain mantles, the deuterium fractionation increases toward the ice surface, which would lead to lower deuterium fractionation ratios in the warm inner regions of protostars compared to their cold outer regions. A good agreement is found between the observations and the predicted deuteration of water and COMs, but it fails to reproduce the high deuteration of the multiply deuterated formaldehyde and methanol. 

To the best of our knowledge, none of the literature studies focused on investigating the best chemical pathway to form deuterated HCOOCH$_3$ among the proposed gas-phase and surface reactions, but most of them test a specific route and/or reaction mechanism. In this study, we focus on the formation of deuterated HCOOCH$_3$ in low-mass star forming regions using a theoretical approach. We test both gas-phase and grain surface chemistries to evaluate their efficiency in forming these isotopologues.

The description of the chemical model used for this study is given in \S \ref{model}. We present and discuss our results and findings in \S \ref{res}. Finally we summarise our main conclusions and remarks in \S \ref{conc}. 
%%%%%%%%%%%%%%%%%%%%%%%%%%%%%%%%%%%%%%%%%%%%%%%%%%%%%
\section{The Model}
\label{model}

\subsection{The physical and chemical model}

In order to understand which chemical pathway is most responsible for the formation of the different forms of deuterated methyl formate (DMF) in star forming regions, we used the open source publicly available astrochemical code UCLCHEM \citep{hold17}\footnote{link to UCLCHEM: https://uclchem.github.io/uclchem.html}. UCLCHEM is a time-dependent, gas-grain chemical model which computes the fractional abundances of the species (with respect to the total number of H in all forms) by solving a set of rate equations. UCLCHEM has been used in previous studies of COMs in star forming regions (e.g. \citealt{cout18, quen18}). 
However, we note that for simplicity, we neglect the temperature dependent diffusion treatment of Langmuir-Hinshelwood reactions that is standard for UCLCHEM \citep{quen18} and instead use grain reactions with fixed rates. Beside the reactions we discuss below, most of the surface reactions included in this work involve mainly simple hydrogenation reactions, with hydrogen moving freely on the grains; hence the temperature dependence is not very important. Most of the non-hydrogenation reactions we include in the network are barrier-less (see Table \ref{tab:reac}) so, again, our simplification seems justified. This allows us to investigate whether a path is viable without considering a large number of unknowns such as binding energy and energy barriers.

Briefly, the fractional abundances are computed in two successive phases; a collapsing phase (Ph I) followed by a warming-up phase (Ph II) which is representative of the hot cores we wish to model. In Ph I, we model the evolution of a diffuse atomic gas which undergoes a free-fall collapse (following \citealt{raw92}) until it reaches a final density suitable for a hot core at a constant temperature of 10 K. During this phase, the chemistry takes place both in the gas-phase and on grain surfaces with active non-thermal desorption processes as described in \citet{robj07}. As a result, this phase provides initial fractional abundances for Ph II that are realistic and consistent with the chemical network rather than assumed.

Once the gas reaches its final density, the second phase of the model starts in order to simulate the effect of the radiation emitted by the new star on the chemical evolution of the dense gas. This radiation gradually heats up its surrounding medium, based on its mass, and causes the thermal desorption of mantle species in a temperature-(hence time) dependent manner according to the experimental work by \citet{col03b}. For this reason, Ph II is known as the warming up phase \citep{viti04}. In this work, as in \citet{viti04} and in \citet{awad10}, the dust temperature increases as a function of time according to the following equation:
\begin{equation}
T_{dust} = 10.0 + (A \times t^B)
\end{equation}
where the constant and the exponent were empirically fitted for a 1 solar mass star (see \citealt{awad10} for full details). Sublimation of mantle species influences the evolution of the gas-phase chemistry during this phase. We note the exact temperature evolution of this phase can be important for temperature dependent grain chemistry \citep{gar13}. However, as noted previously, we do not believe our surface reaction rates depend strongly on the temperature in this case. The initial chemical abundances of this stage (Ph II) are those obtained from the final time step of the collapsing phase (Ph I). 

In the present work, the basic gas-phase chemical network is taken from the KIDA 2014 rate-file\footnote{KIDA website: \url{http://kida.obs.u-bordeaux1.fr}} \citep{wak15}; updated to include the MF formation route proposed by \citet{bal15}; Eq. (\ref{eq:4}). We generated the deuterium network from this gas-phase network following the recipe described in \citet{aik12} and \citet{fur13}. Nuclear spin states of H$_2$, H$_3^+$, and their deuterated isotopologues are included and taken from \citet{hin14}. The grain chemical network relies on simple hydrogenation and deuteration of depleted species, during the collapsing phase, in addition to simple two-body reactions to account for the formation of MF and its deuterated counterparts; see Eqs. (\ref{eq:2} \& \ref{eq:3}). The binding energy of HCOOCH$_3$ was taken from the recent experimental work by \citet{burk15} while we revised all of the desorption energy of our species set from \citet{wak17}. For all the models performed in this study, the chemical network features a total of 8772 reactions, both in the gas-phase and on grain surfaces, linking 252 species (52 of which are mantle species). Differences between the models lie in the active formation route of DMF, but not on the physical conditions which we hold constant for all the performed models.

The species in the network used for this study include all deuterated isotopes of a given species up to the third level (i.e. 3 D atoms per molecule). The initial elemental abundance of atomic D is set to 10$^{-5}$ to reflect the standard interstellar D/H ratio \citep{oli03}. 
The formation of HD on the grains is assumed occur at the same rate as H$_2$. However, \citet{kris11} found that the abundance of D$_2$ molecules on the surface is an order of magnitude higher than HD. Thus, we assume D$_2$ forms at a rate tens times that of HD and H$_2$.
Moreover, we adopt the statistical value of molecular hydrogen ortho-to-para ratio (H$_2$ OPR) to be 3.

Table \ref{tab:initial} lists the initial chemical elemental abundances and physical parameters used for the performed grid of models. The description of the grid of models we ran is given below in \S \ref{grid}.
%%%%%%%%%%%%%%%%%%%%%%%%%%%%%%%%%%%%%%%%%%%%%%%%%%
%%                 Table 1                      %%
%%%%%%%%%%%%%%%%%%%%%%%%%%%%%%%%%%%%%%%%%%%%%%%%%%
\begin{table}
   \centering
\caption{Initial elemental abundances with respect to the total H content (taken from \citealt{asp09}), and physical conditions adopted in our models.}
  \label{tab:initial}
    \leavevmode
    \begin{tabular}{llcll} \hline \hline
\multicolumn{2}{c}{\bf Initial abundances} &&\multicolumn{2}{c}{\bf Physical parameters} \\ \hline
%%%%%%%%%%%%%%%%%%%%%%%%%%%%%%%%%%%%%%%%%%
Helium & 8.50 $\times$ 10$^{-2}$ && Initial density (cm$^{-3}$) & 400 \\
Carbon & 2.69 $\times$ 10$^{-4}$ && Initial temperature (K) & 10 \\
Oxygen & 4.90 $\times$ 10$^{-4}$ && Final density (cm$^{-3}$) & $10^{8}$ \\
Nitrogen & 6.76 $\times$ 10$^{-5}$ && Final temperature (K) & 100 \\
Chlorine & 3.16 $\times$ 10$^{-7}$ && Core radius (AU) & 150 \\
Fluorine & 3.63 $\times$ 10$^{-8}$ && Core Mass (M$_{\odot}$)& 1\\ 
HD & 1.50 $\times$ 10$^{-5}$ && $^{\dag}$Depletion & full \\\hline \hline
\end{tabular}
\flushleft
$\dag$ By full depletion we mean that by the end of collapsing phase more than 90\% of the gas is accreted onto grain surfaces.\\
\end{table}
%%%%%%%%%%%%%%%%%%%%%%%%%%%%%%%%%%%%%%%%%%%%%%%%%%%%%%%%%%%%%%%%%%%%%%%%%%%%%%%%%%%%%%%%%%%%%%%%%%%%%%%%%%%%%%%%%%%%

\subsection{The grid of models}
\label{grid}
For the purpose of this work, we ran eight models, in addition to the reference model ({\bf RM}), to investigate which formation routes of DMF are capable of reproducing the observed abundances in the warm inner regions of low-mass star forming regions. Apart from {\bf RM}, we name each model as ``M+number'' where the number refers to the active routes of DMF formation and corresponds to those listed in Table \ref{tab:reac}.

The reference model ({\bf RM}), in this study, is the model in which none of the new included routes (Eqs. \ref{eq:2}, \ref{eq:3} \& \ref{eq:4}) are active and hence it is somehow chemically equivalent to the model described in \citet{awad14}. In model {\bf M1}, the formation of MF and DMF is active through the surface two-body reaction (CH$_3$OH + CO; Eq. \ref{eq:3}) from \citet{occ11}. For model {\bf M2} the formation of these species occurs through the surface radical-radical reaction (CH$_3$O + HCO; Eq. \ref{eq:2}) from \citet{gar06} while they are formed by the gas-phase scheme (Eq. \ref{eq:4}; taken from \citealt{bal15}) in model {\bf M3}.
These three different routes of formation in the three models are denoted {\bf R1}, {\bf R2} \& {\bf R3}, respectively, in Tables \ref{tab:reac} and \ref{tab:reac-m3}. In reality, the formation of species often takes place through a combination of more than one route. For this reason, we performed a set of 3 models in which each model combines more than one route to form MF and its DMF. These models are model {\bf M12} in which the formation involves only surface reactions (pathways R1 and R2), and models {\bf M13} and {\bf M23} that combine gas- and solid-phase routes. A model in which the three proposed formation schemes are activated is model {\bf M123}.  

Finally, we ran model {\bf Mex} which includes some simple enrichment mechanism such as the H-D exchange (substitution) reactions (listed in Table \ref{tab:reac-ex}) that have been proven, experimentally, to be important in enhancing the deuteration of interstellar species (e.g. \citealt{nag05, hid09, oba16}). The lack of the inclusion of these reactions in chemical models led to a deficiency in the relative abundances of HCOOCH$_2$D and HCOOCHD$_2$ with respect to MF as claimed by \citet{man19}. 

Table \ref{tab:reac} lists all the models we ran and their associated MF and DMF formation pathways. For the combined models, we indicate the routes included by showing their label indicated in column 2 of the table; e.g. for model {\bf M12} the activated routes for formation are R1 and R2. 
%%%%%%%%%%%%%%%%%%%%%%%%%%%%%%%%%%%%%%%%%%%%%%%%%%
%%      Table 2: The main model grid            %%
%%%%%%%%%%%%%%%%%%%%%%%%%%%%%%%%%%%%%%%%%%%%%%%%%%
\begin{table*}
\centering
\caption{Summary of the main grid of models, the new pathways included into our chemical network to form MF, and their rate coefficients ($\alpha, ~ \beta, \&~ \gamma$). The rate constants ($k$, in cm$^{3}$ s$^{-1}$) are calculated using the UMIST formula $k = \alpha~(T/300)^{\beta}~exp(-\gamma/T)$ as described in \citet{mce13}. These reactions are used to generate the pathways of the formation of the different forms of DMF and their rate constants are calculated statistically following \citet{aik12}.}
\label{tab:reac}
\begin{tabular} {lllllllllll} % 6 col
\hline \hline
&&&\multicolumn{4}{c}{\bf Single Pathway Models}&&&\\ %[0.8ex]
%\hline
%---------------------------------------------------------------------------------
{\bf Model}&{\bf Route} &{\bf Re1$^1$} & {\bf Re2$^1$} & & {\bf P1$^1$} & {\bf P2$^1$} & {\bf $\alpha$} & {\bf $\beta$}& {\bf $\gamma$}& {\bf Remarks} \\ \hline %\hline
%---------------------------------------------------------------------------------
%------------------------------------ neutral neutral (surface) ---------------------------------------------
{\bf M1}&{\bf R1} &mCH$_3$OH & mCO & $\longrightarrow$ &  mHCOOCH$_3$ &  & 6.20$\times$10$^{-18}$ & 0 & 0 & \citealt{occ11}\\ [0.8ex]
&& mCH$_3$OD & mCO & $\longrightarrow$ &  mDCOOCH$_3$  &  & 1.55$\times$10$^{-18}$ & 0 & 0 & This work\\[0.8ex]
&& mCH$_3$OD & mCO & $\longrightarrow$ &  mHCOOCH$_2$D &  & 4.65$\times$10$^{-18}$ & 0 & 0 & \\[0.8ex]
&&mCH$_2$DOH & mCO & $\longrightarrow$ &  mDCOOCH$_3$ &  & 1.55$\times$10$^{-18}$ & 0 & 0 & \\[0.8ex]
&&mCH$_2$DOH & mCO & $\longrightarrow$ &  mHCOOCH$_2$D&  & 4.65$\times$10$^{-18}$ & 0 & 0 & \\ [0.8ex]
&&mCH$_2$DOD & mCO & $\longrightarrow$ &  mDCOOCH$_2$D &  & 3.10$\times$10$^{-18}$ & 0 & 0 & \\ [0.8ex]
&&mCH$_2$DOD & mCO & $\longrightarrow$ &  mHCOOCHD$_2$ &  & 3.10$\times$10$^{-18}$ & 0 & 0 & \\[0.8ex]
&&mCHD$_2$OH & mCO & $\longrightarrow$ &  mHCOOCHD$_2$ &  & 3.10$\times$10$^{-18}$ & 0 & 0 & \\ [0.8ex]
&&mCHD$_2$OH & mCO & $\longrightarrow$ &  mDCOOCH$_2$D &  & 3.10$\times$10$^{-18}$ & 0 & 0 & \\\hline
%---------------------------------------------------------------------------------
{\bf M2}&{\bf R2} &mCH$_3$O & mHCO & $\longrightarrow$ &  mHCOOCH$_3$ &  & 1 & 0 & 0 & \citealt{gar08}\\ [0.8ex]
&&mCH$_3$O & mDCO & $\longrightarrow$ &  mDCOOCH$_3$ &  & 1 & 0 & 0 & This work \\[0.8ex]
&&mCH$_2$DO & mHCO & $\longrightarrow$ &  mHCOOCH$_2$D &  & 1 & 0 & 0 & \\[0.8ex]
&&mCH$_2$DO & mDCO & $\longrightarrow$ &  mDCOOCH$_2$D &  & 1 & 0 & 0 & \\[0.8ex]
&&mCHD$_2$O & mHCO & $\longrightarrow$ &  mHCOOCHD$_2$ &  & 1 & 0 & 0 & \\ \hline
%---------------------------------------------------------------------------------
{\bf M3$^{2}$}&{\bf R3} &CH$_3$OH & OH & $\longrightarrow$ & CH$_3$O & H$_2$O & 3.60$\times$10$^{-12}$& -1 & 0 & \citealt{anti16} \\[0.8ex]
&&CH$_3$O  & CH$_3$  & $\longrightarrow$ & CH$_3$OCH$_3$ & photon & 3$\times$10$^{-10}$& 0 & 0 & \citealt{bal15}\\[0.8ex]
&&CH$_3$OCH$_3$  & Cl & $\longrightarrow$ & CH$_3$OCH$_2$ & HCl & 2$\times$10$^{-10}$ & 0 & 0 & \citealt{bal15}\\[0.8ex]
&&CH$_3$OCH$_3$ & F & $\longrightarrow$ & CH$_3$OCH$_2$ & HF & 2$\times$10$^{-10}$ & 0 & 0 & \citealt{bal15}\\[0.8ex]
&&CH$_3$OCH$_2$ & O & $\longrightarrow$ & HCOOCH$_3$ & H & 2$\times$10$^{-10}$& 0 & 0 & \citealt{bal15}\\[0.8ex] 
& &\multicolumn{6}{l}{\bf The deuterated reactions are listed in Table \ref{tab:reac-m3}.} \\
\hline
%-------------------------------
&&&\multicolumn{4}{c}{\bf Multi-pathways Models}&&&\\ [0.8ex] 
%-------------------------------
{\bf M12} &{\bf R1 + R2}&\multicolumn{9}{l}{The formation of MF and DMF is dominated by surface routes.} \\ [0.8ex]
{\bf M13}&{\bf R1 + R3}&\multicolumn{9}{l}{The model combines gas and surface formation routes.}\\ [0.8ex] 
{\bf M23}&{\bf R2 + R3}&\multicolumn{9}{l}{The model combines gas and surface formation routes.} \\ [0.8ex] 
{\bf M123}&{\bf R1 + R2 + R3}&\multicolumn{9}{l}{The model represents the full set of formation pathways on grain surfaces and in the gas-phase.} \\ [0.8ex] 
\hline 
{\bf Mex} &{\bf Rex}&\multicolumn{8}{l}{The set of reactions we added to our chemical network is listed in Table \ref{tab:reac-ex} }& \citealt{oba16} \\ [0.8ex]
\hline \hline
%-------------------------------------------
\end{tabular} 
\flushleft
$^{1}$ The prefix {\bf `Re, P, \& m'} refers to reactants, products, \& mantle species, respectively.\\
$^{2}$ For the {\bf `CH$_3$OH + OH'} reaction, the values of the rate coefficients ($\alpha, ~ \beta, ~ \gamma$) represent the case T $\le$ 100 K. When T $>$ 100 K, these values are 2.85$\times$10$^{-12}$, 0, and 345, respectively \citep{atkin06}. \\
We adopt the statistical branching ratio DCOOCH$_3$:HCOOCH$_2$D = 1:3 \citep{hin14}.\\
\end{table*}
%------------------------------------------
%%%%%%%%%%%%%%%%%%%%%%%%%%%%%%%%%%%%%%%%%%%%%%%%%%
%%    Table 3: Deuterated reactions of route R3 %%
%%%%%%%%%%%%%%%%%%%%%%%%%%%%%%%%%%%%%%%%%%%%%%%%%%
%------------------------------------------
\begin{table*}
\centering
\caption{A list of the deuterated pathways of the reaction set of R3 used in the network of Model M3 as given in Table \ref{tab:reac}.}
\label{tab:reac-m3}
\begin{tabular} {clllllllll} 
\hline \hline
%---------------------------------------------------------------------------------
&&\multicolumn{7}{c}{\bf Single Pathway Models - Deuteration of R3 in Table \ref{tab:reac}}\\ %[0.8ex]
{\bf Model} &{\bf $^1$Re1} & {\bf Re2} & & {\bf P1} & {\bf P2} & {\bf $\alpha$} & {\bf $\beta$}& {\bf $\gamma$}& {\bf Remarks} \\ \hline %\hline
%---------------------------------------------------------------------------------
%---------------------------------------- Balucani (gas-phase)-----------------------------------------
{\bf $^{2}$M3} &CH$_3$OH & OH & $\longrightarrow$ & CH$_3$O & H$_2$O & 3.60$\times$10$^{-12}$& -1 & 0 & T $\le$ 100 K, \citealt{anti16} \\[0.8ex]
&CH$_3$OH  & OD & $\longrightarrow$ & CH$_3$O  & HDO & 3.60$\times$10$^{-12}$& -1 & 0 & This work\\[0.8ex] 
&CH$_3$OD  & OH & $\longrightarrow$ & CH$_3$O  & HDO & 3.60$\times$10$^{-12}$& -1 & 0 &  \\[0.8ex] 
&CH$_3$OD  & OD & $\longrightarrow$ & CH$_3$O  & D$_2$O & 3.60$\times$10$^{-12}$& -1 & 0 & \\[0.8ex] 
&CH$_2$DOH & OH & $\longrightarrow$ & CH$_2$DO & H$_2$O & 3.60$\times$10$^{-12}$& -1 & 0 & \\[0.8ex] 
&CH$_2$DOH & OD & $\longrightarrow$ & CH$_2$DO & HDO & 3.60$\times$10$^{-12}$& -1 & 0 & \\[0.8ex] 
&CHD$_2$OH & OH & $\longrightarrow$ & CHD$_2$O & H$_2$O & 3.60$\times$10$^{-12}$& -1 & 0 & \\[0.8ex] 
&CHD$_2$OH & OD & $\longrightarrow$ & CHD$_2$O & HDO & 3.60$\times$10$^{-12}$& -1 & 0 & \\[0.8ex] 
&CH$_2$DOD & OH & $\longrightarrow$ & CH$_2$DO & HDO & 3.60$\times$10$^{-12}$& -1 & 0 & \\[0.8ex] 
&CH$_2$DOD & OD & $\longrightarrow$ & CH$_2$DO & D$_2$O & 3.60$\times$10$^{-12}$& -1 & 0 & \\%[1ex] 
\\
%%%%%%%%%%%%%%%%%%%%%%%%%%%%%%%%%%%%%%%%%%
&CH$_3$O  & CH$_3$  & $\longrightarrow$ & CH$_3$OCH$_3$ & photon & 3$\times$10$^{-10}$& 0 & 0 & \citealt{bal15}\\[0.8ex]
&CH$_2$DO & CH$_3$  & $\longrightarrow$ & CH$_2$DOCH$_3$ & photon & 3$\times$10$^{-10}$& 0 & 0 & This work \\[0.8ex]
&CH$_3$O  & CH$_2$D & $\longrightarrow$ & CH$_2$DOCH$_3$ & photon & 3$\times$10$^{-10}$& 0 & 0 &  \\[0.8ex]
&CH$_2$DO & CH$_2$D & $\longrightarrow$ & CH$_2$DOCH$_2$D & photon & 3$\times$10$^{-10}$& 0 & 0 &  \\%[1ex]
\\
%%%%%%%%%%%%%%%%%%%%%%%%%%%%%%%%%%%%
&CH$_3$OCH$_3$  & Cl & $\longrightarrow$ & CH$_3$OCH$_2$ & HCl & 2$\times$10$^{-10}$ & 0 & 0 & \citealt{bal15}\\[0.8ex]
&CH$_2$DOCH$_3$ & Cl & $\longrightarrow$ & CH$_2$DOCH$_2$ & HCl & 1$\times$10$^{-10}$ & 0 & 0 & This Work \\[0.8ex]
&CH$_2$DOCH$_3$ & Cl & $\longrightarrow$ & CH$_3$OCH$_2$ & DCl & 3.33$\times$10$^{-11}$ & 0 & 0 &  \\[0.8ex]
&CH$_2$DOCH$_3$ & Cl & $\longrightarrow$ & CH$_3$OCHD & HCl & 6.67$\times$10$^{-11}$ & 0 & 0 &  \\[0.8ex]
&CH$_2$DOCH$_2$D& Cl & $\longrightarrow$ & CH$_2$DOCHD & HCl & 1.33$\times$10$^{-10}$ & 0 & 0 &  \\[0.8ex]
&CH$_2$DOCH$_2$D& Cl & $\longrightarrow$ & CH$_2$DOCH$_2$ & DCl & 6.67$\times$10$^{-11}$ & 0 & 0 &  \\%[1ex]
\\
%%%%%%%%%%%%%%%%%%%%%%%%%%%%%%%%%%%%
&CH$_3$OCH$_3$ & F & $\longrightarrow$ & CH$_3$OCH$_2$ & HF & 2$\times$10$^{-10}$ & 0 & 0 & \citealt{bal15}\\[0.8ex]
&CH$_2$DOCH$_3$ & F & $\longrightarrow$ & CH$_2$DOCH$_2$ & HF & 1$\times$10$^{-10}$ & 0 & 0 & This Work \\[0.8ex]
&CH$_2$DOCH$_3$ & F & $\longrightarrow$ & CH$_3$OCH$_2$ & DF & 3.33$\times$10$^{-11}$ & 0 & 0 & \\[0.8ex]
&CH$_2$DOCH$_3$ & F & $\longrightarrow$ & CH$_3$OCHD & HF & 6.67$\times$10$^{-11}$ & 0 & 0 &  \\[0.8ex]
&CH$_2$DOCH$_2$D& F & $\longrightarrow$ & CH$_2$DOCHD & HF & 1.33$\times$10$^{-10}$ & 0 & 0 &  \\[0.8ex]
&CH$_2$DOCH$_2$D& F & $\longrightarrow$ & CH$_2$DOCH$_2$ & DF & 6.67$\times$10$^{-11}$ & 0 & 0 &  \\%[1ex]
\\
%\newpage
%%%%%%%%%%%%%%%%%%%%%%%%%%%%%%%%%%%
&CH$_3$OCH$_2$ & O & $\longrightarrow$ & HCOOCH$_3$ & H & 2$\times$10$^{-10}$& 0 & 0 & \citealt{bal15}\\[0.8ex] 
&CH$_2$DOCH$_2$ & O & $\longrightarrow$ & HCOOCH$_2$D & H & 2$\times$10$^{-10}$& 0 & 0 & This Work \\[0.8ex] 
&CH$_3$OCHD & O & $\longrightarrow$ & DCOOCH$_3$ & H & 1$\times$10$^{-10}$& 0 & 0 & \\[0.8ex] 
&CH$_3$OCHD & O & $\longrightarrow$ & HCOOCH$_3$ & D & 1$\times$10$^{-10}$& 0 & 0 & \\[0.8ex] 
&CH$_2$DOCHD & O & $\longrightarrow$ & DCOOCH$_2$D & H & 1$\times$10$^{-10}$& 0 & 0 & \\[0.8ex] 
&CH$_2$DOCHD & O & $\longrightarrow$ & HCOOCH$_2$D & D & 1$\times$10$^{-10}$& 0 & 0 & \\%[1ex] 
\hline \hline
%%%%%%%%%%%%%%%%%%%%%%%%%%%%%%%%%%%
\end{tabular} 
\flushleft
$^{1}$ The prefix {\bf `Re, P, \& m'} refers to reactants, products, \& mantle species, respectively.\\
$^{2}$ For the {\bf `CH$_3$OH + OH'} reaction and its deuterated pathways at T $>$ 100 K, the values of the rate coefficients {\bf $\alpha$, $\beta$,} and {\bf $\gamma$} are 2.85$\times$10$^{-12}$, 0, and 345, respectively, according to \citet{atkin06} .\\
\end{table*}
%-----------------------------------------------------------
%%%%%%%%%%%%%%%%%%%%%%%%%%%%%%%%%%%%%%%%%%%%%%%%%%
%%      Table 4: H-D exchange reactions         %%
%%%%%%%%%%%%%%%%%%%%%%%%%%%%%%%%%%%%%%%%%%%%%%%%%%
%------------------------------------------
\begin{table*}
\centering
\caption{List of exchange reactions we added to our network for model Mex following \citet{oba16}.}
\label{tab:reac-ex}
\begin{tabular} {lllllllllll} % 6 col
\hline \hline
{\bf Model}&&{\bf $^1$Re1} & {\bf Re2} & & {\bf P1} & {\bf P2} & {\bf $\alpha$} & {\bf $\beta$}& {\bf $\gamma$}& {\bf Remarks} \\ \hline 
{\bf Mex} && mHCOOCH$_3$ & mHDO & $\longrightarrow$ & mHCOOCH$_2$D & mH$_2$O & 0.5 & 0 & 0 \\ [0.8 ex]
&& mHCOOCH$_3$ & mHDO &  $\longrightarrow$ & mDCOOCH$_3$ & mH$_2$O & 0.5 & 0 & 0 \\ [0.8 ex]
&& mHCOOCH$_3$ & mD$_2$O &  $\longrightarrow$ & mHCOOCH$_2$D & mHDO & 0.5 & 0 & 0 \\ [0.8 ex]
&& mHCOOCH$_3$ & mD$_2$O &  $\longrightarrow$ & mDCOOCH$_3$ & mHDO & 0.5 & 0 & 0 \\ [0.8 ex]
&& mHCOOCH$_2$D & mH$_2$O &  $\longrightarrow$ & mHCOOCH$_3$ & mHDO & 1 & 0 & 0 \\ [0.8 ex]
&& mHCOOCH$_2$D & mHDO &  $\longrightarrow$ & mHCOOCH$_3$ & mD$_2$O & 0.2 & 0 & 0 \\ [0.8 ex]
&& mHCOOCH$_2$D & mHDO &  $\longrightarrow$ & mHCOOCHD$_2$ & mH$_2$O & 0.6 & 0 & 0 \\ [0.8 ex]
&& mHCOOCH$_2$D & mHDO &  $\longrightarrow$ & mDCOOCH$_2$D & mH$_2$O & 0.2 & 0 & 0 \\ [0.8 ex]
&& mHCOOCH$_2$D & mD$_2$O &  $\longrightarrow$ & mHCOOCHD$_2$ & mHDO & 0.75 & 0 & 0 \\ [0.8 ex]
&& mHCOOCH$_2$D & mD$_2$O &  $\longrightarrow$ & mDCOOCH$_2$D & mHDO & 0.25 & 0 & 0 \\ [0.8 ex]
&& mHCOOCHD$_2$ & mH$_2$O &  $\longrightarrow$ & mHCOOCH$_2$D & mHDO & 1 & 0 & 0 \\ [0.8 ex]
&& mHCOOCHD$_2$ & mHDO &  $\longrightarrow$ & mHCOOCD3 & mH$_2$O & 0.2 & 0 & 0 \\ [0.8 ex]
&& mHCOOCHD$_2$ & mHDO &  $\longrightarrow$ & mDCOOCHD$_2$ & mH$_2$O & 0.2 & 0 & 0 \\ [0.8 ex]
&& mHCOOCHD$_2$ & mHDO &  $\longrightarrow$ & mHCOOCH$_2$D & mD$_2$O & 0.6 & 0 & 0 \\ [0.8 ex]
&& mHCOOCD3 & mH$_2$O &  $\longrightarrow$ & mHCOOCHD$_2$ & mHDO & 1 & 0 & 0 \\ [0.8 ex]
&& mHCOOCD3 & mHDO &  $\longrightarrow$ & mHCOOCHD$_2$ & mD$_2$O & 1 & 0 & 0 \\ [0.8 ex]
&& mDCOOCH$_3$ & mH$_2$O &  $\longrightarrow$ & mHCOOCH$_3$ & mHDO & 1 & 0 & 0 \\ [0.8 ex]
&& mDCOOCH$_3$ & mHDO &  $\longrightarrow$ & mHCOOCH$_3$ & mD$_2$O & 0.25 & 0 & 0 \\ [0.8 ex]
&& mDCOOCH$_3$ & mHDO &  $\longrightarrow$ & mDCOOCH$_2$D & mH$_2$O & 0.75 & 0 & 0 \\ [0.8 ex]
&& mDCOOCH$_3$ & mD$_2$O &  $\longrightarrow$ & mDCOOCH$_2$D & mHDO & 1 & 0 & 0 \\ [0.8 ex]
&& mDCOOCH$_2$D & mH$_2$O &  $\longrightarrow$ & mHCOOCH$_2$D & mHDO & 0.25 & 0 & 0 \\ [0.8 ex]
&& mDCOOCH$_2$D & mH$_2$O &  $\longrightarrow$ & mDCOOCH$_3$ & mHDO & 0.75 & 0 & 0 \\ [0.8 ex]
&& mDCOOCH$_2$D & mHDO &  $\longrightarrow$ & mHCOOCH$_2$D & mD$_2$O & 0.2 & 0 & 0 \\ [0.8 ex]
&& mDCOOCH$_2$D & mHDO &  $\longrightarrow$ & mDCOOCH$_3$ & mD$_2$O & 0.4 & 0 & 0 \\ [0.8 ex]
&& mDCOOCH$_2$D & mHDO &  $\longrightarrow$ & mDCOOCHD$_2$ & mH$_2$O & 0.4 & 0 & 0 \\ [0.8 ex]
&& mDCOOCH$_2$D & mD$_2$O &  $\longrightarrow$ & mDCOOCHD$_2$ & mHDO & 1 & 0 & 0 \\ [0.8 ex]
&& mDCOOCHD$_2$ & mH$_2$O &  $\longrightarrow$ & mHCOOCHD$_2$ & mHDO & 0.25 & 0 & 0 \\ [0.8 ex]
&& mDCOOCHD$_2$ & mH$_2$O &  $\longrightarrow$ & mDCOOCH$_2$D & mHDO & 0.75 & 0 & 0 \\ [0.8 ex]
&& mDCOOCHD$_2$ & mHDO &  $\longrightarrow$ & mHCOOCHD$_2$ & mD$_2$O & 0.25 & 0 & 0 \\ [0.8 ex]
&& mDCOOCHD$_2$ & mHDO &  $\longrightarrow$ & mDCOOCH$_2$D & mD$_2$O & 0.75 & 0 & 0 \\ [0.8 ex]
\hline \hline
%---------------------------------------------------------------------------------
\end{tabular} 
\flushleft
$^{1}$ The prefix {\bf `Re, P, \& m'} refers to reactants, products, \& mantle species, respectively.\\
\end{table*}
%%%%%%%%%%%%%%%%%%%%%%%%%%%%%%%%%%%%%%%%%%%%%%%%%%%%%%%%%%%%%%%%%%%%%%%%%%%%%%%%%%%%%%%%%%%%%%%%%%%%%%%%%%%%%%%%%%

\section{Results and Discussion}
\label{res}
The time evolution of the fractional abundances during the warming up phase (Ph II) of methyl formate (MF) and its deuterated isotopologues (DMF) are shown in Figs. \ref{fig:1} to \ref{fig:3} under the physical conditions of a typical 1 M$_{\odot}$ hot corino. Different curves in each figure indicate different chemical models (see figure keys and Table \ref{tab:reac}). In all the plots, different models are displayed with different line-styles where the reference model ({\bf RM}) is denoted by dash--dot lines; thick straight lines, parallel to the x--axis, represent the upper limits of the abundances of MF and DMF derived for IRAS~16293~B (see Sect. \ref{comp}). In addition, if a species peaks above 1 $\times$ 10$^{-12}$ in abundance and has been observed, we limit y--axis to 1 $\times$ 10$^{-12}$, otherwise the lower limit of the y--axis is set to 1 $\times$ 10$^{-20}$. 
Tables \ref{tab:obs} and \ref{tab:ratio} summarise the predicted abundances of MF and its different forms and the associated D/H ratios during the warming up phase (Ph II) at time $t$ = 1 -- 2 $\times$ 10$^5$ years. For comparison, the observed values in the protostellar object IRAS 16293--2422 are also indicated.

\subsection{Sensitivity to formation pathways}
\label{network}
\begin{figure*}
\begin{center}
% trim left bottom right top)
\includegraphics[trim=0.5cm 0.3cm 0.3cm 0.3cm, clip=true, width=16cm]{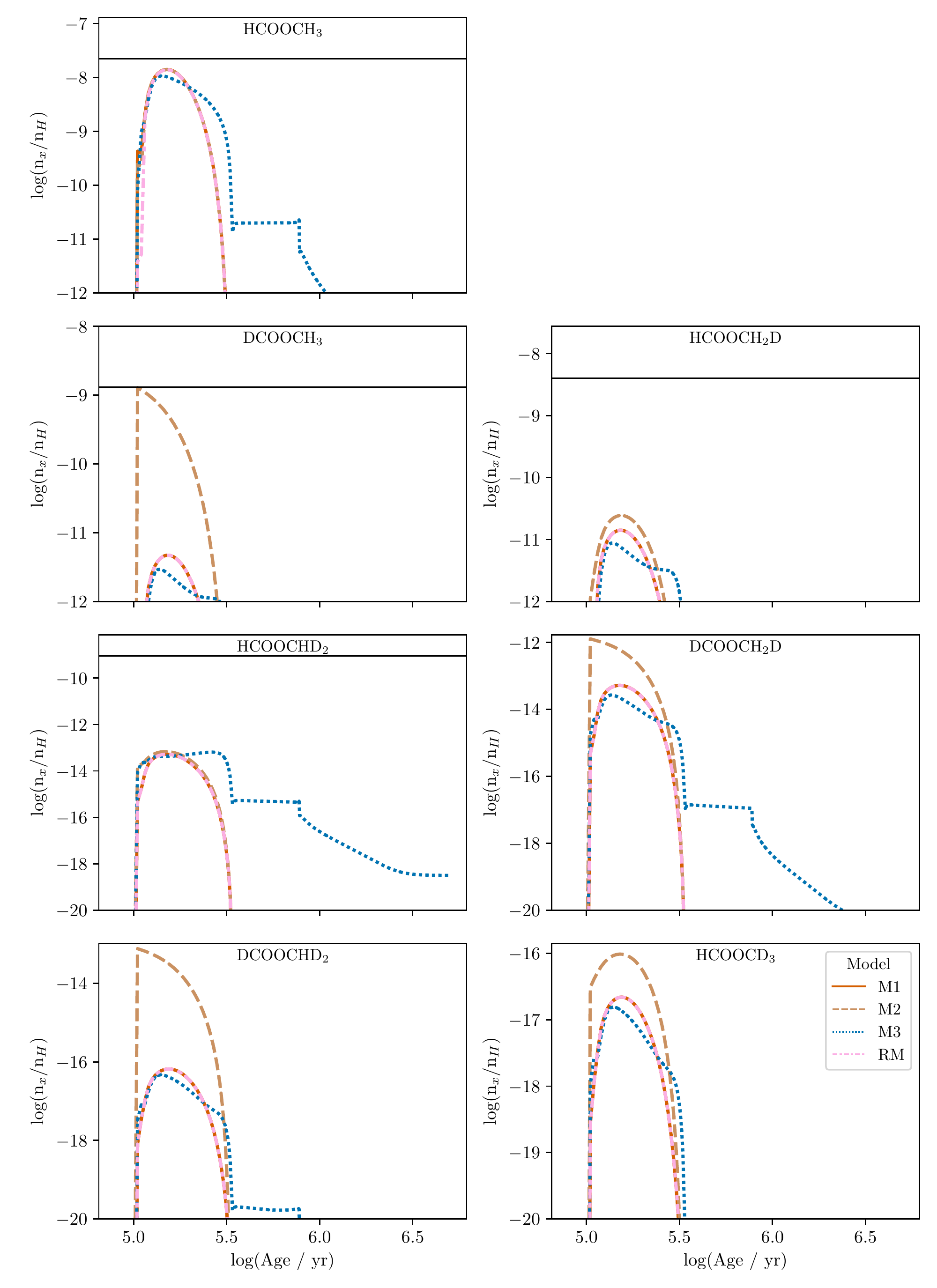}
\caption {The time evolution of the computed fractional abundances of MF and DMF in models in which we test the effect of including a single formation pathway to the network either on grain surfaces, {\bf M1} and {\bf M2}, or in the gas-phase, {\bf M3}. The results are compared to the {\bf RM}, dash--dot line, where neither of the three routes are included; see figure key. Straight grey lines represent the upper limits of the MF and DMF abundances in IRAS 16293B \citep{man19,jor16}; see Table \ref{tab:obs}.}
\label{fig:1}
\end{center}
\end{figure*}

As illustrated in the figures, the obtained evolutionary trends of the species, from all performed models, show similarities. All the species are detectable at times around 10$^5$ yrs that correspond to the time of their thermal evaporation from grains. At these times, the fractional abundances increase until they reach a maximum value. After that, the abundances decline until they become undetectable at times around (3 -- 4) $\times$ 10$^5$ yrs. None of our single pathway models ({\bf M1}, {\bf M2}, and {\bf M3}) shown in Fig. \ref{fig:1} nor the {\bf RM} is able to form HCOOCD$_3$ with detectable abundances.

Model {\bf M2}, dash lines in Fig. \ref{fig:1}, produces significantly more DCOOCH$_3$ and DCOOCH$_2$D (1 -- 2 orders of magnitude), and it is capable of forming very small amounts of HCOOCHD$_2$ and DCOOCHD$_2$ than the other models. This shows the efficiency of radical-radical reactions (route R2) in the formation of isotopologues of COMs. 

In model {\bf M3}, where the formation of MF and DMF occur via gas-phase chemistry, we found that the abundances of MF and both of its mono-deuterated forms, DCOOCH$_3$ and HCOOCH$_2$D, remain in the gas for longer times compared to the other models, and the abundance of MF remain relatively high for times up to $\sim$ 2.5 $\times$ 10$^6$ yrs. The analysis of HCOOCH$_3$ , DCOOCH$_3$, and HCOOCH$_2$D showed that, at these longer times, the species are mainly formed in the gas-phase via oxidisation reactions of CH$_3$OCH$_2$, CH$_3$OCHD, and CH$_2$DOCH$_2$, respectively. It also revealed that the main source of destruction of MF and DMF are the HCO$^+$ and DCO$^+$ ions which compete with H$_3$O$^+$ up to  3.2 $\times$ 10$^5$ yrs, after which the abundance of H$_3$O$^+$ becomes around 30 times higher than them, and hence it dominates the destruction of these species. For the time interval up to $\sim$ 2 $\times$ 10$^5$ yrs, the formation rate of the three species is 10--20 times higher than that of the destruction, and therefore, we see an increase in their abundance. After that and until 3.2 $\times$ 10$^5$ yrs, the formation and destruction rates of the three species become comparable and hence we see the plateau feature in the plots. When the abundance of H$_3$O$^+$ becomes higher than that of the main parents, at 3.2 $\times$ 10$^5$ yrs, the destruction rate increases leading to a decline of the molecular abundances until they disappear from the medium at a time of $\sim$ 4 $\times$ 10$^5$ (for mono-DMFs) or 2.5 $\times$ 10$^6$ yrs (for MF).
%%%%%%%%%%%%%%%%%%%%%%%%%%%%%%%%%%%%%%%%%%%%%%%%%%
%%                 Fig 1                        %%
%%%%%%%%%%%%%%%%%%%%%%%%%%%%%%%%%%%%%%%%%%%%%%%%%%
%%%%%%%%%%%%%%%%%%%%%%%%%%%%%%%%%%%%%%%%%%%%%%%%%%%%%%%%%%%%%%%%%%%%%%%%%%%%%%%%%%%%%%%%%%%%%%%%%%%%

\begin{figure*}
\begin{center}
% trim left bottom right top
\includegraphics[trim=0.5cm 0.3cm 0.3cm 0.3cm, clip=true,width=16cm]{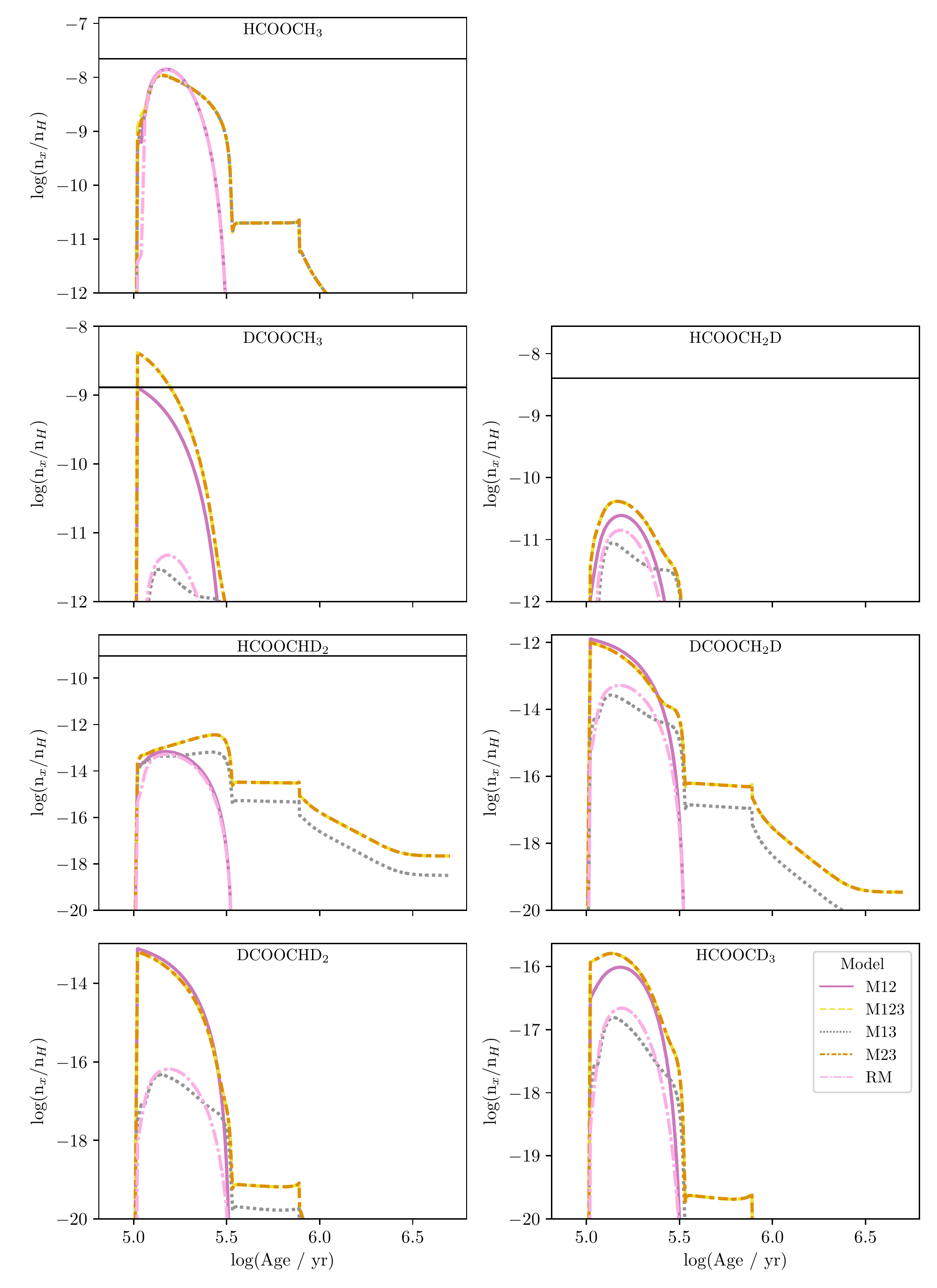}
\caption {Similar to Fig. \ref{fig:1}, but for models that examine combining more than one formation pathway; see Table \ref{tab:reac}. The results are also compared to the RM denoted by dash--dot line.}
\label{fig:2}
\end{center}
\end{figure*}

We defined the multi-pathway models to be those in which the chemical network includes more than one formation route of MF and DMF; see Table \ref{tab:reac}. Model calculations are represented in Fig. \ref{fig:2} together with those of the RM. The results show that the peak abundances of most of the species are not significantly affected by combining more than one formation route. Models {\bf M12}, {\bf M23}, and {\bf M123} (where route R2 is common) show the highest abundances.
%%%%%%%%%%%%%%%%%%%%%%%%%%%%%%%%%%%%%%%%%%%%%%%%%%
%%                 Fig 2                        %%
%%%%%%%%%%%%%%%%%%%%%%%%%%%%%%%%%%%%%%%%%%%%%%%%%%
%%%%%%%%%%%%%%%%%%%%%%%%%%%%%%%%%%%%%%%%%%%%%%%%%%%%%%%%%%%%%%%%%%%%%%%%%%%%%%%%%%%%%%%%%%%%%%%%%%%%
These findings support the results by \citet{gar08} in which the authors showed that radical-radical reactions may dominate the formation process of complex molecules on grain surfaces in dense interstellar regions. 
Indeed, the chemical analysis of DCOOCH$_3$ and DCOOCH$_2$D in models {\bf M12}, {\bf M23}, and {\bf M123} revealed that the surface radical-radical route (R2) dominates their formation pathways during both phases Ph I and Ph II. During the cold collapsing phase (Ph I), where deuteration may occur efficiently, the analysis showed that the formation rates of these species in the three models are at least 100 times higher than their destruction rates, especially towards the end of the collapse. During these late times, the density of the core becomes more than 10$^6$ cm$^{-3}$, which weakens the non-thermal evaporation processes of mantle species. This would allow longer residence time for the parent molecules of these species on grain surfaces and hence would increase their yield. As a consequence, when DCOOCH$_3$ and DCOOCH$_2$D evaporate from grain mantles in the warming up phase, they enrich the medium with higher abundances compared to other models. 

\begin{figure*}
\begin{center}
% trim left bottom right top
\includegraphics[trim=0.0cm 0.3cm 0.3cm 0.3cm, clip=true, width=16cm]{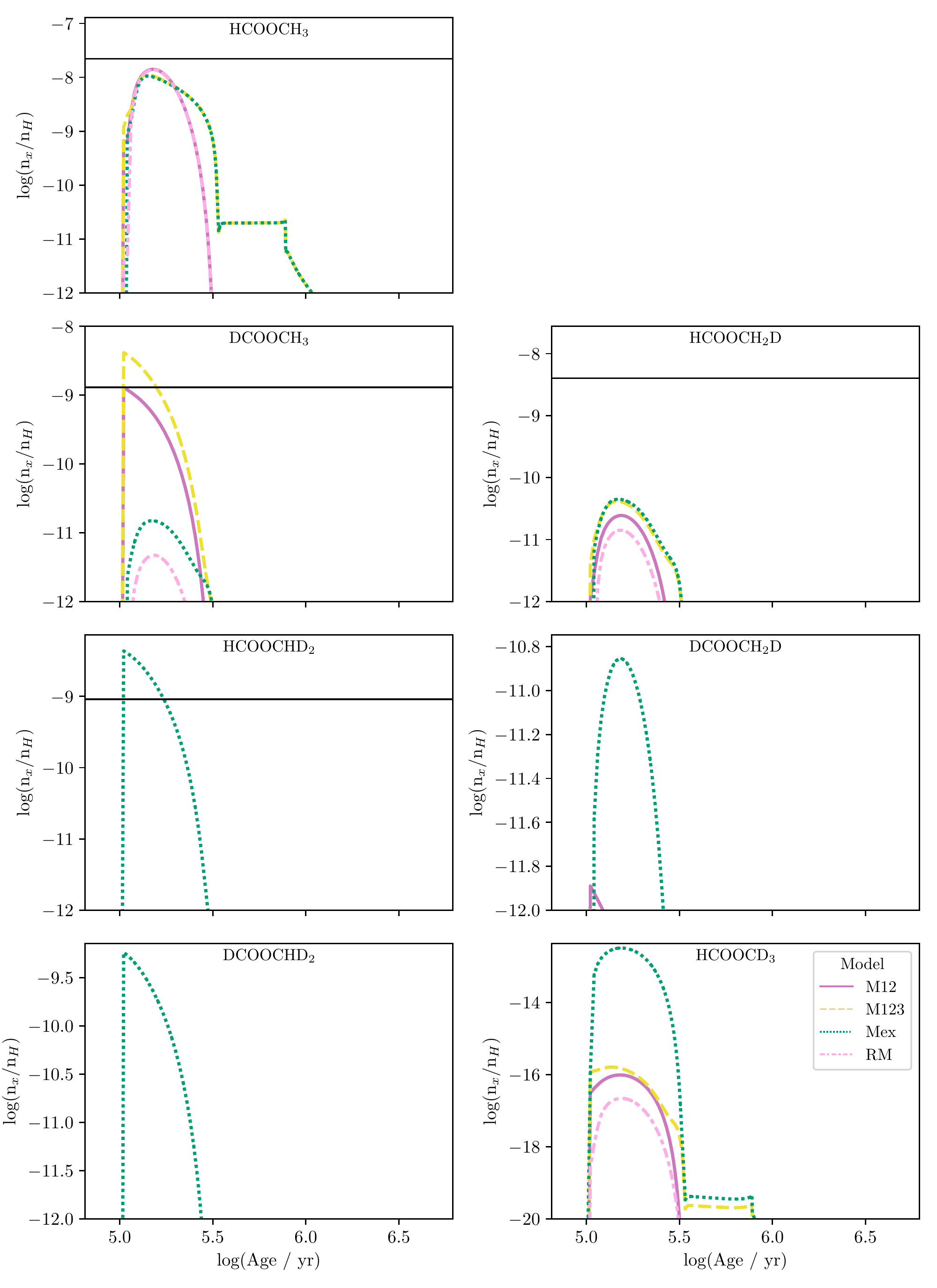}
\caption {The chemical evolution of MF and DMF in hot corinos assuming that only H-D substitution/exchange reactions are their formation routes. The results are compared to the abundances of these species in models where the formation is dominated by surface chemistry, {\bf M12}, or models that reflect the mutual gas-grain interaction pathways, {\bf M123}; see figure key.}
\label{fig:3}
\end{center}
\end{figure*}

The results of model, {\bf Mex}, with those from {\bf M12} (only surface pathways) and {\bf M123} (full gas-grain pathways), are illustrated in Fig. \ref{fig:3}. We observe for {\bf Mex} an increase in the molecular abundances of the forms of multiple DMF (DCOOCH$_2$D, HCOOCHD$_2$, DCOOCHD$_2$ and HCOOCD$_3$) and a significant decrease (at the time of evaporation) in the abundance of DCOOCH$_3$ (2 -- 3 orders of magnitude). HCOOCH$_2$D is insensitive to the inclusion of the H-D exchange reactions. 
The chemical analysis of the results of model {\bf Mex} revealed that DCOOCH$_3$ forms DCOOCH$_2$D which acts as the parent molecule of DCOOCHD$_2$. This sequence shows that the deuteration occurs for the CH$_3$ group. In addition, the same analysis was obtained for HCOOCD$_3$ which is the daughter of HCOOCHD$_2$. These results are inline with those obtained experimentally by \citet{nag05}. The authors reported that the H-D substitution occur in the CH$_3$ group of methanol (CH$_3$OH) and not on the OH group and that is why they did not observe any molecules on the form Me-d$_n$-OD, where Me-d$_n$ represents the deuteration in CH$_3$ group with n = 0 -- 3. Our model shows that, in addition to radical-radical reactions on grain surfaces, H-D substitution reactions can play a major role in the enrichment of the medium with deuterated complex molecules.

%%%%%%%%%%%%%%%%%%%%%%%%%%%%%%%%%%%%%%%%%%%%%%%%%%
%%%                 Fig 3                       %%
%%%%%%%%%%%%%%%%%%%%%%%%%%%%%%%%%%%%%%%%%%%%%%%%%%
%%%%%%%%%%%%%%%%%%%%%%%%%%%%%%%%%%%%%%%%%%%%%%%%%%%%%%%%%%%%%%%%%%%%%%%%%%%%%%%%%%%%%%%%%%%%%%%%%%%%

\subsection{Comparison with observations}
\label{comp}

This work focuses on the formation pathways of the deuterated isotopologues of MF under physical and chemical conditions comparable to those observed in hot corinos, in particular the one in the Class 0 protostar IRAS 16293--2422 ({\it hereafter} IRAS 16293). In this section, we are going to discuss our model calculations in light of the most recent ALMA PILS observations of MF and DMF in IRAS 16293 \citep{jor18, man19}. We may also shed light on these species in other low-mass protostellar objects such as NGC1333 IRAS4A \citep{bot04b} and SVS13-A \citep{bia19}.

The fractional abundances, $x$(X), obtained with our chemical models are expressed with respect to the total hydrogen density of the core, n$_H$. In order to compare our results with observations, we converted the observed column densities of the species, N(X), into fractional abundances with respect to the total hydrogen density. For IRAS 16293B, \citet{jor16} derived an H$_2$ column density N(H$_2$) of $\gtrsim$ 1.2 $\times$ 10$^{25}$ cm$^{-2}$. We computed the fractional abundances of MF and DMF in IRAS 16293B by dividing the column densities determined by \citet{jor18} and \citet{man19} by the lower limit of N(H$_2$). The abundance is multiplied by 2 to take into account the fact that the position analysed in these studies is shifted by one beam offset with respect to the continuum peak position \citep{jor16,jor18}. These fractional abundances are then computed with respect to the total H following a simple expression n(X)/n$_{\text H}$ = 0.5 n(X)/n$_{\text {H$_2$}}$ \citep{pety05}.

The abundance of HCOOCH$_3$ (with respect to H) is estimated to be $\lesssim$ 2.2 $\times$ 10$^{-8}$ in IRAS 16293 \citep{jor18,man19}. All of our models predict abundances in agreement with the observations ($\sim$ 1.4 $\times$ 10$^{-8}$ during the warming up phase at $t$ $\sim$ 1 -- 2 $\times$ 10$^{5}$ years). These results are also inline with the observed abundances in NGC1333 IRAS4A (4 $\times$ 10$^{-8}$; \citealt{bot04b}) and in the more chemically evolved Class I protostar SVS13-A (2.2 $\times$ 10$^{-8}$; \citealt{bia19}).

None of the computed models can reproduce  the observed abundances of MF and all the detected forms of DMF, at once, during the warming-up phase of the hot corino. All the models reproduce well the observed abundances of MF, but only the models including R2 ({\bf M2}, {\bf M12}, {\bf M23} and {\bf M123}) and {\bf Mex} are able to reproduce the observed amounts of DCOOCH$_3$ and HCOOCHD$_2$, respectively. In addition, all of our models predict abundances of HCOOCH$_2$D that are about 100 times lower than the observations. More details can be found below for the different isotopologues (see also Tables  \ref{tab:obs} and \ref{tab:ratio}).

DCOOCH$_3$ was detected in IRAS 16293 with an abundance of $\lesssim$ 1.3 $\times$ 10$^{-9}$ and a D/H ratio of 1.7\% in IRAS 16293A and 5.7\% in IRAS 16293B \citep{jor18,man19}. Models {\bf M2}, {\bf M12}, {\bf M23} and {\bf M123} that includes a formation through radical-radical association on grains (R2), reproduce well these observations, once the species evaporate from grains at 1.1 $\times$ 10$^{5}$ years. These models predict abundances of 9 $\times$ 10$^{-10}$ -- 5 $\times$ 10$^{-9}$ at a time of $\sim$ 1 -- 2 $\times$ 10$^{5}$ years (when MF best fit observations).

The DCOOCH$_3$/HCOOCH$_3$ abundance ratios predicted by Models {\bf M2} and {\bf M12} ($\sim$ 6.3\%) is in perfect agreement with the value measured towards IRAS 16293B (5.7\%) and only 3.7 times higher than the ratio determined in IRAS 16293A \citep{jor18, man19} while the ratios estimated from the other models ({\bf M23} \& {\bf M123}; 35\%) are about 6 times larger than observations for IRAS 16293B. These results indicate the importance of radical-radical reactions, route R2 in Table \ref{tab:reac}, in the formation of this molecule. It also implies that R2 dominates the other chemical routes of formation of DCOOCH$_3$ as reflected by the comparable abundances predicted in the four models that best fit the observations (Models {\bf M2}, {\bf M12}, {\bf M23} and {\bf M123}). It is worth noting that the DCOOCH$_3$/HCOOCH$_3$ ratios for models {\bf M2} and {\bf M12} are inline with the predictions of the so-called early-time model ($\sim$ 9.2\% at 1.1 $\times$ 10$^{5}$ years) by \citet{taq14}. 

The predicted abundances of HCOOCH$_2$D and HCOOCHD$_2$ are found to be lower than the observations by a factor of 10$^{2}$ and 10$^3$, except for the model {\bf Mex} for HCOOCHD$_2$ which is discussed later. 
HCOOCH$_2$D is detected with an abundance of $\lesssim$ 4 $\times$ 10$^{-9}$, while the models predict abundances of $\le$ 4 $\times$ 10$^{-11}$. 
Similarly for HCOOCHD$_2$, which has an observed abundance of $\lesssim$ 9 $\times$ 10$^{-10}$, while the models (except for Mex) 
predict no more than 4 $\times$ 10$^{-13}$. The abundances ratios of HCOOCH$_2$D with respect to MF are 8\% and 18\% and those of HCOOCHD$_2$ 
are 2\% and 4\%, in IRAS~16293 A and B, respectively. The equivalent ratios predicted by our models are significantly lower (0.3\% for HCOOCH$_2$D and 0.003\% for HCOOCHD$_2$), which would mean that some deuteration reaction pathways are missing in our network. 
In comparison, \citet{taq14} predicted for HCOOCH$_2$D an abundance ratio with respect to MF of 21\% at a time of 1.1 $\times$ 10$^5$ yrs but much smaller (0.6\%) and comparable to our results at a time of 2 $\times$ 10$^5$ yrs.
Based on these results, the radical-radical reactions included in our models are not sufficient to explain the observed abundances of these two molecules.

When introducing the H-D exchange reactions (Table \ref{tab:reac-ex}, Model {\bf Mex}), we see a surprising enhancement of the abundance of HCOOCHD$_2$ ($\sim$ 1.7 $\times$ 10$^{-9}$ at 1.5 $\times$ 10$^5$ years) which is about 4 orders of magnitude higher than the results of the other models in the grid. The predicted abundance is twice the observed value in IRAS 16293B ($\lesssim$ 9 $\times$ 10$^{-10}$), but it decreases with time and at a time of 2.5 $\times$ 10$^5$ years, a good agreement is obtained with respect to the observed value. As a consequence, we obtain a high HCOOCHD$_2$/MF abundance ratio of 12\% at a time of 1 -- 2 $\times$ 10$^5$ years. The H-D substitution reactions can explain the high abundance of HCOOCHD$_2$ but they are however not sufficient to reproduce the observed abundance of HCOOCH$_2$D.

In summary, the comparison of our models with the observations of IRAS 16293 shows that the formation of the different forms of DMF relies on both radical-radical associations and H-D exchange reactions on grains. 
Surface reactions involving radicals (R2; \citealt{gar08}) are sufficient to explain the observed abundance of DCOOCH$_3$, but H-D exchange reactions are also necessary to reproduce the abundance of the doubly deuterated form HCOOCHD$_2$ in IRAS 16293. It should be noted that H-D exchange reactions would also imply high abundances of DCOOCHD$_2$, which could consequently be detectable. To the best of our knowledge, this isotopologue is however not available in the spectroscopic databases. The deuteration pathways of MF may be more complex than expected given the difficulty in reproducing the observations of HCOOCH$_2$D. The chemical network may be missing some effective formation pathways or it may be that temperature dependent diffusion is indeed needed to achieve a  higher abundance of HCOOCH$_2$D.
%%%%%%%%%%%%%%%%%%%%%%%%%%%%%%%%%%%%%%%%%%%%%%%%%%
%%     Table 5: comparison abundances           %%
%%%%%%%%%%%%%%%%%%%%%%%%%%%%%%%%%%%%%%%%%%%%%%%%%%
\begin{table*}
\centering
\caption{Comparison between the observed fractional abundances of MF and DMF in the protostellar object IRAS 16293B and those obtained with our models after the evaporation of the grain ices (t $\sim$ 1 -- 2 $\times$ 10$^5$ years) in Ph-II. a(b) means a$\times$10$^b$.}
\label{tab:obs}
\leavevmode
\begin{tabular}{llll} \hline \hline
{\bf Species} & {\bf $^{\dag}$IRAS 16293B} & {\bf This work} & {\bf Corresponding Model}\\ [1.0 ex] \hline
% --------------------------------------------------------------------------
{\bf $^{(1)}$HCOOCH$_3$}  & $\lesssim$ 2.2 (-8) & 1.4 (-8)& All \\ [1.0 ex] 
% --------------------------------------------------------------------------
{\bf $^{(1)}$DCOOCH$_3$} & $\lesssim$ 1.3 (-9) & 9 (-10) & M2, M12\\ [1.0 ex] 
                        &                     & 5 (-9) &  M23, M123 \\ [1.0 ex]
% -------------------------------------------------------------------------
{\bf $^{(1)}$HCOOCH$_2$D} & $\lesssim$ 4.0 (-9)&  $\le$ 4.0 (-11) & All \\ [1.0 ex] 
%-------------------------------------------------------------------------
{\bf $^{(2)}$HCOOCHD$_2$} & $\lesssim$ 9.1 (-10) & $\le$ 4.0 (-13) & All except Mex \\ [1.0 ex]
                         &           & 1.7 (-9)    & Mex \\ [1.0 ex] 
% --------------------------------------------------------------------------
{\bf $^{(3)}$CH$_3$OH} & $\lesssim$ 8.3 (-7) & 2.3 (-6) & RM, M1, M2, M12\\ 
                     &                      & 3.0 (-7) & Mex, M3, M13, M23, M123\\\hline \hline
% ----------------------------------------------------------------------------
\end{tabular} 
\flushleft
$^{\dag}$ Observations are converted into fractional abundances with respect to the total number of H as described in \S (\ref{comp}).\\
{\bf References:} (1) \citealt{jor18}, (2) \citealt{man19}, (3) \citealt{jor16}.
\end{table*}
%%%%%%%%%%%%%%%%%%%%%%%%%%%%%%%%%%%%%%%%%%%%%%%%%%%%%%%%%%%%%%%%%%%%%%%%%%%%%%%%%%%%%
%%%%%%%%%%%%%%%%%%%%%%%%%%%%%%%%%%%%%%%%%%%%%%%%%%
%%      Table 6: comparison ratios              %%
%%%%%%%%%%%%%%%%%%%%%%%%%%%%%%%%%%%%%%%%%%%%%%%%%%
\begin{table*}
\centering
\caption{Comparison between the DMF-to-MF column density ratios in IRAS 16293 and the theoretically calculated ratios during Ph-II at a time of 1 -- 2 $\times$ 10$^5$ yrs. The last two columns of the table list the ratios obtained with the model by \citet{taq14}. a(b) means a$\times$10$^b$.}
\label{tab:ratio}
\leavevmode
\begin{tabular}{llllllll} \hline \hline
{\bf Abundance ratios} & \multicolumn{2}{c}{\bf IRAS 16293} && {\bf This Work} &{\bf Corresponding Model} & \multicolumn{2}{c}{\bf Other Model $^{(3)}$}\\ [0.8 ex]
   & {\bf A $^{(1)}$}  &  {\bf B $^{(1,2)}$} &&    &  & Early time & Late time \\ \hline
% --------------------------------------------------------------------------
{\bf DCOOCH$_3$/HCOOCH$_3$} & 1.7\% & 5.7\% && 6.2\% & M2, M12 & 9.2\% & 0.0025\% \\ [1.0 ex] %\hline
% --------------------------------------------------------------------------
{\bf HCOOCH$_2$D/HCOOCH$_3$} & 8.5\% & 18.0\% &&  0.3\% &  All & 21\% & 0.65\% \\ [1.0 ex] %\hline
% --------------------------------------------------------------------------
{\bf $^{\dag}$HCOOCHD$_2$/HCOOCH$_3$} & 2.0\% & 4.2\% && 0.003\% & All except Mex & 2.4\% & 0.0016\% \\ [1.0 ex] 
                                    &       &       && 12\% &  Mex & \\ [1.0 ex] %\hline
% --------------------------------------------------------------------------
{\bf $^{\ddag}$HCOOCH$_3$/CH$_3$OH} & 2.1\% & 2.7\%  && 0.6\% & RM, M1, M2, M12 & --- & --- \\ [1.0 ex] 
                                  &       &        && 4.7\% & Mex, M3, M13, M23, M123 & --- & --- \\ [1.0 ex]
\hline \hline
% ---------------------------------------------------------------------------
\end{tabular} %\\
\flushleft
$^{\dag}$ Taken from Table 2 in \citet{man19}\\
$^{\ddag}$ The observed ratio in IRAS 16293A is taken from \citet{man20} while for IRAS 16293B, we computed the ratio by adopting N(HCOOCH$_3$) and N(CH$_3$OH) observations from \citet{man19} and \citet{jor16}, respectively. \\
{\bf References:} (1) \citealt{man19}, (2) \citealt{jor18}, (3) \citealt{taq14}\\
\end{table*}
%%%%%%%%%%%%%%%%%%%%%%%%%%%%%%%%%%%%%%%%%%%%%%%%%%

%%%%%%%%%%%%%%%%%%%%%%%%%%%%%%
\section{Conclusions}
\label{conc}
This work examined the formation of deuterated methyl formate (DMF) in light of the proposed chemical pathways 
to form methyl formate (MF) both in the gas-phase and on grain surfaces in low-mass star forming regions. 

The main conclusion here is that the formation of all forms of DMF cannot be explained by a single scenario, at once, because species 
respond differently to changes in the chemistry. It is clear therefore that any future astrochemical modelling ought to include a more rigorous surface network where, not only all the reactions explored here need to be included, but also the temperature dependence of the diffusion treatment of the Langmuir-Hinshelwood reactions, as well as surface reactions during the warm-up phase, should be included.  
Our models succeed in reproducing the observed abundances of HCOOCH$_3$, DCOOCH$_3$ and 
HCOOCHD$_2$ after their evaporation from grains at times of 1 -- 2 $\times$ 10$^5$ years. However, they always underestimate the abundances of 
HCOOCH$_2$D by a factor of $\sim$ 100. Our results show that radical-radical reactions often dominate over other chemical pathways.
Models with H-D substitution reactions are capable of producing doubly and multiply deuterated MF such as DCOOCHD$_2$ and HCOOCD$_3$. 
In addition, we found that the deuteration of the OH group is more sensitive to changes in the chemistry than the one of the CH$_3$ group. 
The calculations also showed that the abundance ratio between D(CH$_3$) and D(OH) deviates from the initially assumed statistical value of 3.

Our models predict the existence of multiply deuterated isotopologues of methyl formate, DCOOCH$_2$D, DCOOCHD$_2$ and HCOOCD$_3$ in low-mass star 
forming regions with abundances up to an order of 10$^{-10}$, 10$^{-9}$ and 10$^{-12}$, respectively, assuming that substitution reactions are 
enhancing the molecular abundances as suggested experimentally (e.g. \citealt{nag05, hid09, oba16}). 

It is clear that the chemistry of DMF is more complicated than the simple picture we presented here.  Comprehensive work, experimentally as well as theoretically, is needed for a better understanding of the interstellar chemistry of these species, especially reactions on grain surfaces. 

%%%%%%%%%%%%%%%%%%%%%%%%%%%%%%%%%%%%%%%%%%%%%%
\section*{Acknowledgements}
\label{ack}
ZA would like to thank Dr. Ugo Hincelin for providing her with his basic deuterium chemical network that was modified, updated and partially used in this work. AC acknowledges financial support from the Agence Nationale de la Recherche (grant ANR-19-ERC7-0001-01). JH and SV acknowledge funding  by the European Research Council (ERC) Advanced Grant MOPPEX 833460. We thank the anonymous referee for their constructive comments.
%%%%%%%%%%%%%%%%%%%%%%%%%%%%%%%%%%%%%%%%%%%%%%%%%
\section*{Data AVAILABILITY}
The model outputs underlying this article will be freely shared on request to the corresponding author.
%%%%%%%%%%%%%%%%%%%%%%%%%%%%%%%%%%%%%%%%%%%%%%%%%%
%%                 REFERENCES                   %%
%\bibliographystyle{mnras}
%\bibliography{D-references2020} % if your bibtex file is called example.bib
%%%%%%%%%%%%%%%%%%%%%%%%%%%%%%%%%%%%%%%%%%%%%%%%%%

%%%%%%%%%%%%%%%%% APPENDICES %%%%%%%%%%%%%%%%%%%%%
%\newpage
%\section*{Appendix}
%\label{apn}
%%%%%%%%%%%%%%%%%%%%%%%%%%%%%%%%%%%%%%%%

% Don't change these lines
\bsp % typesetting comment
\label{lastpage}
\end{document}